\documentclass[acus]{JAC2001}
\usepackage{graphicx}
\newcommand{\ud}{\mathrm{d}}
\begin{document}
\onecolumn 
\thispagestyle{empty}

\begin{center}
\begin{tabular}{p{130mm}}

\begin{center}
{\bf\Large
MULTISCALE DECOMPOSITION FOR} \\
\vspace{5mm}

{\bf\Large VLASOV-POISSON EQUATIONS}

\vspace{1cm}

{\bf\Large Antonina N. Fedorova, Michael G. Zeitlin}\\

\vspace{1cm}

{\bf\large\it
IPME RAS, St.~Petersburg, 
V.O. Bolshoj pr., 61, 199178, Russia}\\
{\bf\large\it e-mail: zeitlin@math.ipme.ru}\\
{\bf\large\it http://www.ipme.ru/zeitlin.html}\\
{\bf\large\it http://www.ipme.nw.ru/zeitlin.html}
\end{center}

\vspace{1cm}

\abstract{
We consider the applications of a nu\-me\-ri\-cal\--analytical 
approach based on multiscale variational wavelet technique
to the
systems with collective type behaviour described by 
some forms of Vlasov-Poisson/Maxwell equations.
We calculate the exact fast convergent representations for solutions in
high-localized
wavelet-like bases functions,
which correspond to underlying hidden (coherent) nonlinear eigenmodes.
This helps to control
stability/unstability  scenario of evolution in parameter 
space on pure algebraical level.
} 

\vspace{60mm}

\begin{center}
{\large Presented at the Eighth European Particle Accelerator Conference} \\
{\large EPAC'02} \\
{\large Paris, France,  June 3-7, 2002}
\end{center}
\end{tabular}
\end{center}
\newpage

\title{MULTISCALE DECOMPOSITION FOR VLASOV-POISSON EQUATIONS}
\author{Antonina  N. Fedorova, Michael  G. Zeitlin \\
IPME, RAS, V.O. Bolshoj pr., 61, 199178, St.~Petersburg, Russia
\thanks{e-mail: zeitlin@math.ipme.ru}\thanks{ http://www.ipme.ru/zeitlin.html;
http://www.ipme.nw.ru/zeitlin.html}   
}
\maketitle

\begin{abstract}

We consider the applications of a nu\-me\-ri\-cal\--analytical 
approach based on multiscale variational wavelet technique
to the
systems with collective type behaviour described by 
some forms of Vlasov-Poisson/Maxwell equations.
We calculate the exact fast convergent representations for solutions in
high-localized
wavelet-like bases functions,
which correspond to underlying hidden (coherent) nonlinear eigenmodes.
This helps to control
stability/unstability  scenario of evolution in parameter 
space on pure algebraical level.

\end{abstract}

\section{INTRODUCTION}

In this paper we consider the applications of nu\-me\-ri\-cal\--analytical 
approach based on multiscale variational wavelet technique
to the
systems with collective type behaviour described by 
some forms of Vlasov-Poisson/Maxwell equations [1], [2].
Such approach may be useful in all models in which  it is 
possible and reasonable to reduce all complicated problems related with 
statistical distributions to the problems described 
by the systems of nonlinear ordinary/partial differential/integral 
equations with or without some (functional) constraints.
In periodic accelerators and transport systems at the high beam currents 
and charge densities the effects of the intense self-fields, which are
produced by the beam space charge and currents, 
determinine (possible) equilibrium states,
stability and transport properties according to
underlying nonlinear dynamics [2].
The dynamics of such space-charge dominated high
brightness beam systems
can provide the understanding of
the instability phenomena such as
emittance growth,
mismatch, halo formation related to the
complicated behaviour of underlying hidden nonlinear modes
outside of perturbative tori-like KAM regions.
Our analysis is based on the 
variational-wavelet approach from [3]-[17],
which allows us to consider polynomial and rational type of 
nonlinearities.
In some sense our approach is direct generaliztion of traditional
nonlinear $\delta F$ approach in which 
weighted Klimontovich representation 
\begin{eqnarray}
\delta f_j=a_j\sum^{N_j}_{i=1}w_{ji}\delta(x-x_{ji})\delta(p-p_{ji})
\end{eqnarray}
or self-similar decompostion [2] like
\begin{eqnarray}
\delta n_j=b_j\sum^{N_j}_{i=1}w_{ji} s(x-x_{ji}),
\end{eqnarray}
where
$s(x-x_{ji})$ is a shape function of distributing particles 
on the grids in configuration space,
are replaced by powerful technique from local nonlinear harmonic analysis,
based on underlying symmetries of functional space such as affine or more general.
The solution has the multiscale/multiresolution decomposition via 
nonlinear high-localized eigenmodes, 
which corresponds  to the full multiresolution expansion in all 
underlying time/phase space scales. 
Starting  from Vlasov-Poisson equations in part 2,
we consider the approach based on
multiscale variational-wavelet formulation in part 3. 
We give the explicit representation for all dynamical variables in the base of
compactly supported wavelets or nonlinear eigenmodes. Our solutions
are parametrized
by solutions of a number of reduced algebraical 
problems one from which
is nonlinear with the same degree of nonlinearity 
as initial problem and the others  are
the linear problems which correspond to the particular
method of calculations inside concrete wavelet scheme. 
Because our approach started from variational formulation we can control
evolution of instability on the pure algebraical level of reduced
algebraical system of equations. 
This helps to control
stability/unstability scenario of evolution in parameter 
space on pure algebraical level.
In all these models numerical modeling demonstrates the appearance 
of coherent high-localized structures and
as a result the stable patterns formation or unstable chaotic behaviour.

\section{VLASOV-POISSON EQUATIONS}

Analysis based on the non-linear Vla\-sov 
equations leads to more
clear understanding of  collective effects and nonlinear beam dynamics
of high intensity beam propagation in pe\-ri\-o\-dic\--fo\-cu\-sing 
and uni\-form\--fo\-cu\-sing
transport systems.
We consider the following form of equations (ref. [1] for setup and designation): 
{\setlength\arraycolsep{0pt}
\begin{eqnarray}
&&\Big\{\frac{\partial}{\partial s}+p_x\frac{\partial}{\partial x}+
             p_y\frac{\partial}{\partial y}-
\Big[k_x(s)x+\frac{\partial\psi}{\partial x}\Big]\frac{\partial}{\partial p_x}-\nonumber\\
&& \Big[k_y(s)y+\frac{\partial\psi}{\partial y}\Big]\frac{\partial}{\partial p_y}
  \Big\} f_b(x,y,p_x,p_y,s)=0, \\
&&\Big(\frac{\partial^2}{\partial x^2}+\frac{\partial^2}{\partial y^2}\Big)\psi=
-\frac{2\pi K_b}{N_b}\int \ud p_x \ud p_y f_b,\\
&&\int\ud x\ud y\ud p_x\ud p_y f_b=N_b
\end{eqnarray}}
The corresponding Hamiltonian for transverse sing\-le\--par\-ticle motion is given by 
{\setlength\arraycolsep{0pt}
\begin{eqnarray}
&& H(x,y,p_x,p_y,s)=\frac{1}{2}(p_x^2+p_y^2) 
                   +\frac{1}{2}[k_x(s)x^2 \\
 &&+k_y(s)y^2]+
    H_1(x,y,p_x,p_y,s)+\psi(x,y,s), \nonumber
\end{eqnarray}}
where $H_1$ is nonlinear (polynomial/rational) part of the full Hamiltonian 
and corresponding characteristic equations are:
\begin{eqnarray}
\frac{\ud^2x}{\ud s^2}+k_x(s)x+\frac{\partial}{\partial x}\psi(x,y,s)&=&0\\
\frac{\ud^2y}{\ud s^2}+k_y(s)y+\frac{\partial}{\partial y}\psi(x,y,s)&=&0
\end{eqnarray}

\section{MULTISCALE REPRESENTATIONS}

We obtain our multiscale/multiresolution representations for solutions of equations
(3)-(8) via variational-wavelet approach. We decompose the solutions as 
{\setlength\arraycolsep{0pt} 
\begin{eqnarray}
&&f_b(s,x,y,p_x,p_y)=\sum^\infty_{i=i_c}\oplus\delta^if(s,x,y,p_x,p_y)\\
&&\psi(s,x,y)=\sum^\infty_{j=j_c}\oplus\delta^j\psi(s,x,y)\\
&&x(s)=\sum^\infty_{k=k_c}\oplus\delta^kx(s),\quad
y(s)=\sum^\infty_{\ell=\ell_c}\oplus\delta^\ell y(s)
\end{eqnarray}}
where set $(i_c,j_c,k_c,\ell_c)$ corresponds to the coarsest level of resolution
$c$ in the full multiresolution decomposition
\begin{equation}
V_c\subset V_{c+1}\subset V_{c+2}\subset\dots
\end{equation}
Introducing detail space 
$W_j$ as the orthonormal complement of $V_j$ with respect to $V_{j+1}: 
V_{j+1}=V_j\bigoplus W_j$,
we have for $f$, $\psi$, $x$, $y$ $\subset L^2({\bf R})$ from (9)-(11):
\begin{eqnarray}
L^2({\bf R})=\overline{V_c\displaystyle\bigoplus^\infty_{j=c} W_j},
\end{eqnarray}
In some sense (9)-(11) is some generalization of the old $\delta F$ approach [1], [2].
Let $L$ be an arbitrary (non) \-li\-ne\-ar dif\-fe\-ren\-tial\-/in\-teg\-ral 
operator with matrix dimension $d$, 
which acts on some set of functions
$\Psi\equiv\Psi(s,x)=\Big(\Psi^1(s,x),\dots,\Psi^d(s,x)\Big)$, $ s,x \in\Omega\subset{\bf R}^{n+1}$
from $L^2(\Omega)$:
\begin{equation}
L\Psi\equiv L(R(s,x),s,x)\Psi(s,x)=0,
\end{equation}
($x$ are the generalized space coordinates or phase space coordinates, $s$ is ``time'' coordinate).
After some anzatzes [3]-[17] the main reduced problem may be formulated as the system of ordinary 
differential            
equations                                                              
{\setlength\arraycolsep{0pt}                                                                
\begin{eqnarray}\label{eq:pol0}                                
& & Q_i(f)\frac{\ud f_i}{\ud s}=P_i(f,s),\quad f=(f_1,..., f_n),\\
& &i=1,\dots,n, \quad                                                                        
 \max_i  deg \ P_i=p, \quad \max_i deg \  Q_i=q \nonumber                  
\end{eqnarray}} 
\noindent or a set of such systems corresponding to each independent coordinate
in phase space. 
They have the fixed initial (or boundary) conditions $f_i(0)$, where $P_i, Q_i$ are not more    
than polynomial functions of dynamical variables $f_j$                                 
and  have arbitrary dependence on time. 
As result
we have the following reduced algebraical system
of equations on the set of unknown coefficients $\lambda_i^k$ of
localized eigenmode expansion (formula (17) below):
\begin{eqnarray}\label{eq:pol2}
L(Q_{ij},\lambda,\alpha_I)=M(P_{ij},\lambda,\beta_J),
\end{eqnarray}
where operators L and M are algebraization of RHS and LHS of initial problem
(\ref{eq:pol0}) and $\lambda$ are unknowns of reduced system
of algebraical equations (RSAE) (\ref{eq:pol2}).
After solution of RSAE (\ref{eq:pol2}) we determine
the coefficients of wavelet expansion and therefore
obtain the solution of our initial problem.
It should be noted that if we consider only truncated expansion with N terms
then we have from (\ref{eq:pol2}) the system of $N\times n$ algebraical equations
with degree $\ell=max\{p,q\}$
and the degree of this algebraical system coincides
with degree of initial differential system.
So, we have the solution of the initial nonlinear
(rational) problem  in the form
\begin{eqnarray}\label{eq:pol3}
f_i(s)=f_i(0)+\sum_{k=1}^N\lambda_i^k f_k(s),
\end{eqnarray}
where coefficients $\lambda_i^k$ are the roots of the corresponding
reduced algebraical (polynomial) problem RSAE (\ref{eq:pol2}).
Consequently, we have a parametrization of solution of initial problem
by the solution of reduced algebraical problem (\ref{eq:pol2}).
The obtained solutions are given
in the form (\ref{eq:pol3}),
where
$f_k(t)$ are basis functions obtained via multiresolution expansions (9)-(11), (13) 
and represented by
some compactly supported wavelets.
As a result the solution of equations (3)-(8) has the 
following mul\-ti\-sca\-le\-/mul\-ti\-re\-so\-lu\-ti\-on decomposition via 
nonlinear high\--lo\-ca\-li\-zed eigenmodes, 
which corresponds to the full multiresolution expansion in all underlying  
scales (13) starting from coarsest one
(polynomial tensor bases are introduced in [17]; ${\bf x}=(x,y,p_x,p_y)$):
{\setlength\arraycolsep{1pt}
\begin{eqnarray}\label{eq:z}
\Psi(s,{\bf x})&=&\sum_{(i,j)\in Z^2}a_{ij}{\bf U}^i\otimes V^j(s,{\bf x}),\\
V^j(s)&=&V_N^{j,slow}(s)+\sum_{l\geq N}V^j_l(\omega_ls), \quad \omega_l\sim 2^l \nonumber\\
{\bf U}^i({\bf x})&=&{\bf U}_M^{i,slow}({\bf x})+
\sum_{m\geq M}{\bf U}^i_m(k_m{\bf x}), \quad k_m\sim 2^m, \nonumber
\end{eqnarray}}
\begin{figure}[htb]
\centering
\includegraphics*[width=65mm]{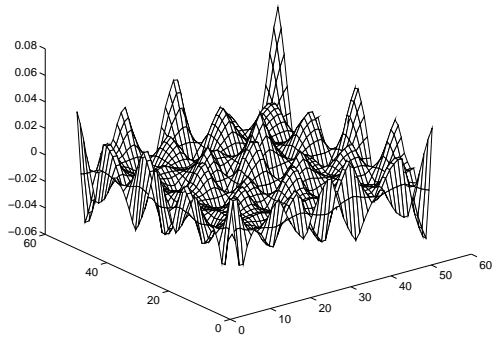}
\caption{Eigenmode of level 1.}
\end{figure}
\begin{figure}[htb]
\centering
\includegraphics*[width=65mm]{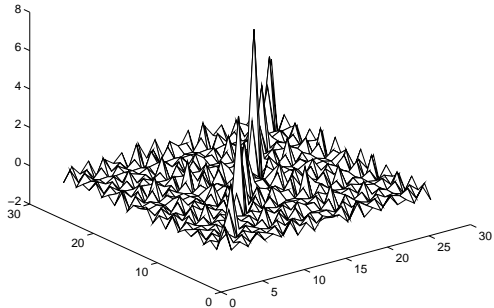}
\caption{Stable waveleton pattern.}
\end{figure}
\begin{figure}[htb]
\centering
\includegraphics*[width=65mm]{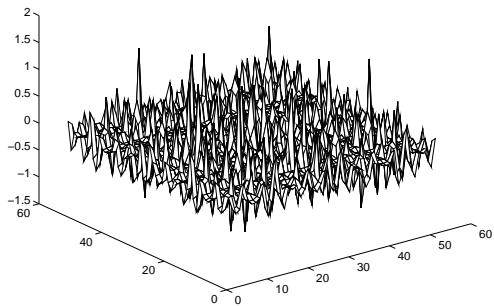}
\caption{Chaotic-like behaviour.}
\end{figure}
Formula (\ref{eq:z}) gives us expansion into the slow part $\Psi_{N,M}^{slow}$
and fast oscillating parts for arbitrary N, M.  So, we may move
from coarse scales of resolution to the 
finest one for obtaining more detailed information about our dynamical process.
The first terms in the RHS of formulae (18) correspond on the global level
of function space decomposition to  resolution space and the second ones
to detail space. 
It should be noted that such representations 
give the best possible localization
properties in the corresponding (phase)space/time coordinates. 
In contrast with different approaches formulae (18) do not use perturbation
technique or linearization procedures.
So, by using wavelet bases with their good (phase) space/time      
localization properties we can describe high-localized (coherent) structures in      
spa\-ti\-al\-ly\--ex\-te\-nd\-ed stochastic systems with collective behaviour.
Modelling demonstrates the appearance of stable patterns formation from
high-localized coherent structures or chaotic behaviour.
On Fig.~1 we present contribution to the full expansion from coarsest level (waveleton) 
of decomposition (18). Fig. 2, 3 show the representations for full solutions, constructed
from the first 6 eigenmodes (6 levels in formula (18)), and demonstrate stable localized 
pattern formation and chaotic-like behaviour outside of KAM region.
We can control the type of behaviour on the level of reduced algebraical system (16).

\end{document}